\documentclass[prb,onecolumn,superscriptaddress,notitlepage,nofootinbib,longbibliography]{revtex4-2}

\usepackage{physics}
\usepackage{amsmath}
\usepackage{amsthm}
\usepackage{amssymb}
\usepackage{bm}
\usepackage{dcolumn}
\usepackage[mathlines]{lineno}
\usepackage{graphicx}
\usepackage{subfigure}
\usepackage[utf8]{inputenc}
\usepackage[T1]{fontenc}
\usepackage[colorlinks,linkcolor=blue,anchorcolor=blue,citecolor=blue]{hyperref}
\begin{document}

\title{Spacetime Foam Effects on Charged AdS Black Hole Thermodynamics} 

\author{Yahya Ladghami}
\email{yahya.ladghami@ump.ac.ma}
\affiliation{Laboratory of Physics of Matter and Radiation, Mohammed I University, BP 717,
Oujda, Morocco}
\affiliation{Astrophysical and Cosmological Center, BP 717, Oujda, Morocco}
\author{Aatifa Bargach}
\email{a.bargach@ump.ac.ma} 
\affiliation{Laboratory of Physics of Matter and Radiation, Mohammed I University, BP 717,
Oujda, Morocco}
\affiliation{Astrophysical and Cosmological Center, BP 717, Oujda, Morocco}
\affiliation{Multidisciplinary Faculty of Nador, University of Mohammed first, BP. 300, Selouane, Nador 62702, Morocco}
\author{Amine Bouali}
\email{a1.bouali@ump.ac.ma}
\affiliation{Laboratory of Physics of Matter and Radiation, Mohammed I University, BP 717,
Oujda, Morocco}
\affiliation{Astrophysical and Cosmological Center, BP 717, Oujda, Morocco}
\affiliation{ Higher School of Education and Training, Mohammed I University, BP 717, Oujda, Morocco}

\author{Taoufik Ouali}
\email{t.ouali@ump.ac.ma}\affiliation{Laboratory of Physics of Matter and Radiation, Mohammed I University, BP 717, Oujda, Morocco}
\affiliation{Astrophysical and Cosmological Center, BP 717, Oujda, Morocco}
\author{G. Mustafa}
\email{gmustafa3828@gmail.com}
\affiliation{Department of Physics, Zhejiang Normal University, Jinhua 321004, China}
\affiliation{Research Center of Astrophysics and Cosmology, Khazar University, Baku, AZ1096, 41 Mehseti Street, Azerbaijan}
\date{\today}

\begin{abstract}
In this paper, we investigate the emergent thermodynamic phenomena arising from spacetime foam and its impact on black hole behavior. Within this framework, we adopt the Barrow model, where the structure of spacetime at small scales is modeled by analogy with the Koch snowflake, implying that black hole surfaces acquire a quasi-fractal structure due to quantum deformations induced by quantum gravity effects. Our analysis, conducted within the extended phase-space formalism, reveals that the quasi-fractal correction to black hole entropy significantly modifies the equation of state, critical parameters, and phase-transition behavior of charged AdS black holes. An increase in the Barrow parameter leads to higher critical pressure and temperature, which diverge at maximal deformation. Moreover, while the quasi-fractal structure has a negligible effect on small black holes with low entropy, it clearly influences the thermal evolution of medium and large event horizon black holes.  Additionally, we study the impact of quasi‑fractal corrections on the Joule–Thomson expansion and the phase transition between cooling and heating regimes. We also examine the effects of spacetime structure on black hole microstate density, lifetimes, and temperature detection by different observers, including local, asymptotic, and Unruh detectors. We find that spacetime foam increases microstate density and prolongs evaporation lifetimes, thus acting as a resistance to black hole evaporation, while local observers experience that the expected Tolman blueshift and Unruh temperatures remain unmodified.

\end{abstract}

\maketitle

\section{Introduction}
The study of black hole thermodynamics has recently attracted considerable interest since it provides profound insights into the nature of black holes and their connection to fundamental principles governing the behavior of physical systems. Indeed, the key concepts of black hole thermodynamics originated with the work of Hawking, Bekenstein, Bardeen, and Carter \cite{Bardeen_1973gs} where they have formulated four laws of black hole mechanics analogous to those of classical thermodynamics. It was also found that black holes radiate thermally \cite{Hawking_1975vcx} and should be assigned an entropy that is proportional to the area of its event horizon \cite{Bekenstein_1973ur}.

Furthermore, the Hawking-Page phase transition \cite{Hawking_1982dh}, introduced by Stephen Hawking and Don Page in 1983, deals with the thermodynamic properties of black holes in anti-de Sitter (AdS) space, describing a transition between the Schwarzschild-AdS black hole and pure thermal AdS space. This is the traditional black hole thermodynamics (TBHT) approach, where important properties of black holes are studied without considering pressure and volume, and the cosmological constant $\Lambda$ is not initially treated as a thermodynamic variable. An extension of this approach, incorporating additional variables, was introduced by Kastor et al. \cite{Kastor_2009wy} to investigate Van der Waals-like phase transitions in charged AdS black holes. This is known as the extended phase space thermodynamics (EPST) formalism \cite{Dolan_2011xt, Kubiznak_2012wp}. This formalism is obtained by including a negative cosmological constant as a thermodynamic variable, i.e., the cosmological constant is related to the pressure through equation
$P = -\Lambda/8\pi G$ where the pressure and its conjugate variable, the thermodynamic volume $V = (\partial M/ \partial P) _{S,Q,J}$, form a new pair of thermodynamic variables (P, V). This investigation gains more physical meaning. Indeed, besides the fact that the phase transition of charged AdS black holes coincides remarkably with the Van der Waals liquid-gas phase transition, various types of black holes in AdS space also exhibit the same phase transitions \cite{Belhaj_2013cva, Li_2014ixn, R.G.Cai, Gunasekaran}. Namely, phase transitions between small and large black holes \cite{Kubiznak}, stretched quintessence phase \cite{AoP}, multiple critical points \cite{Tavakoli}, polymer-type phase transitions \cite{Dolan}, and Joule-Thomson expansion \cite{Okcu_2016tgt} (We will revisit the topic of Joule–Thomson expansion later). A new study of black holes has appeared in the framework of the AdS/CFT correspondence \cite{AdS} since black holes are equivalent to a thermal state in CFT. Known as holographic thermodynamics, this study has attracted a great deal of attention \cite{HT1,HT2,HT3,HT33,HT4,HT5,Ya1,Ya2}.

 another context and  in order to reconcile quantum and gravity concepts, John Wheeler introduced the concept of spacetime foam by proposing that at Planck-length scales spacetime undergoes quantum fluctuations and has a complex structure because of quantum gravitational effects, resulting in a foam‐like geometry \cite{W55,W56,W57}. Later on, Hawking  proposed that this foam consists of virtual black holes \cite{yahya}, and one of his most important predictions was that the Higgs boson could never be detected, but this model was disproved after the Higgs boson was observed \cite{Higgs}. Other works have explored alternative realizations of spacetime foam, including wormholes \cite{w59} and quantum time machine  \cite{w60}. Furthermore, all leading approaches to quantum gravity support the notion of spacetime foam: loop quantum gravity predicts a discretized spacetime that resembles such foam \cite{SF,SF1}, and in string theory the quantum production of D-branes has been interpreted as another manifestation of spacetime foam \cite{SF2,SF3}. 

Recently, in order to advance the study of the small-scale structure of spacetime and its consequences, John Barrow modeled spacetime foam using the Koch snowflake, suggesting that at very small scales spacetime possesses a quasi-fractal structure \cite{Barrow_2020tzx}. Because the Barrow model is based on a phenomenological analogy with the Koch snowflake and does not satisfy all the conditions required for a complete fractal model, such as rigorously defining the fractal dimensions \cite{ABC}, we refer to it as a quasi-fractal.  Although spacetime foam is inherently a quantum phenomenon, it can give rise to emergent, observable effects in gravitational systems. Several studies have investigated the influence of fractal geometry on cosmology, including its impact on Big Bang nucleosynthesis \cite{Barrow_2021BBB}, the age of the Universe \cite{Sheykhi_2023}, cosmic inflation \cite{Luciano_2023roh}, and holographic dark energy \cite{Saridakis_2020}. Other research has shown that fractal spacetime can emerge in quantum gravity without reference to spacetime foam: for example, causal dynamical triangulations \cite{147} and the exact renormalization group  approach \cite{148} both predict a fractal spacetime at the Planck scale. Moreover, quantum field theory and cosmology have been formulated on fractal spacetime using renormalizable field theories \cite{149}, and observational constraints on such fractal cosmology models have been established \cite{150}. Furthermore, the fractal structure of spacetime has been interpreted in the context of q-derivative theories \cite{151}, with corresponding particle physics constraints \cite{152} and cosmological applications \cite{153}. Additionally, there are other theories of multifractal spacetime, such as those involving ordinary derivatives and weighted derivatives, where these theories study the Big Bang, inflation, the expansion of the Universe, the existence of dark energy, and unitarity \cite{153a}. On the other hand, there are quantum field theories with fractional operators, particularly ultraviolet-complete quantum field theories based on fractional powers of the d'Alembertian operator. These approaches extend to fractional gravity theories, where both renormalizability and unitarity are shown to be compatible \cite{153b}.

In our paper, we use the Barrow model to study the impact of spacetime structure on black hole thermodynamics, where spacetime foam is modeled using fractal geometry. In this framework, the event horizon of a black hole is also affected by spacetime foam. According to Barrow's model, the surface of a black hole acquires a fractal structure, which modifies its entropy as follows \cite{Barrow_2020tzx}
\begin{equation}
    S_B = \left(\frac{A}{4G}\right)^{1+\delta/2},
\end{equation}
where \(S_B\) is the Barrow entropy, \(A\) is the surface area of the black hole, \(G\) is the Newton's constant, and \(\delta\) is a Barrow parameter quantifying the quantum deformation of the event horizon due to spacetime foam.

Using the Barrow model, we can explore emergent phenomena and the impact of the quasi-fractal structure of spacetime on black hole behavior. We believe that black hole thermodynamics, particularly for AdS black holes, provides an excellent laboratory for investigating and uncovering emergent phenomena arising from quantum gravitational effects, as it unifies gravitational and quantum effects within a single formalism. These reasons motivate our investigation into the emergent phenomena and the influence of a fractal spacetime structure on the thermodynamic behavior of black holes.

The Joule-Thomson effect investigates how the temperature of a gas alters as it undergoes a pressure-modifying process. Within this mechanical framework, the temperature may either rise or fall. Certain gases exhibit a critical point where a transition occurs between an increasing and a decreasing temperature. Mathematically, it is given by $\mu =(\partial T/ \partial P)_H$ with $\mu$ as the Joule-Thomson coefficient.  The Joule-Thomson process has been studied before for different configurations of AdS black holes and in different gravitational theories (see \cite{Okcu_2017qgo,Mo_2018rgq,Lan_2018nnp, Mo_2018qkt,Cisterna_2018jqg,RizwanCL_2018cyb, MahdavianYekta_2019dwf,Bi_2020vcg, MoraisGraca_2021ife,sk1,sk2,sk3} for more details).

The remainder of the paper is organized as follows: In Sect. \ref{SS1}, we integrate the effects of quantum gravity
on charged black holes asymptotically in AdS spacetime. We calculate the cosmological constant (as it is interpreted as the pressure), the thermodynamic volume, and the Barrow entropy in EPST formalism. In Sect. \ref{SS2},  we explore the influence of spacetime foam on thermodynamic processes, phase transition, and the stability of black holes. Our attention in Sect. \ref{SS3} shifts to exploring the effects of the quasi-fractal structure on the Joule-Thomson expansion. Within this framework, we investigate how the Barrow temperature varies with pressure while maintaining a constant mass for the black hole. In Sect. \ref{ss11}, we study the impact of spacetime foam on microstate density and black hole evaporation.  Finally, in Sect. \ref{SS4}, we present our conclusions.

\section{Spacetime Foam  and Charged AdS Black Holes}
\label{SS1}
In this section, we determine the thermodynamic quantities and correct the Smarr relation by incorporating the foam-like nature of spacetime at small scales. We adopt Barrow's model to simplify the study of the structure of spacetime and its effects on black holes.

Due to the quasi-fractal nature of the black hole surface and in the context of the Bekenstein–Hawking area law, Barrow proposed that the corrected entropy is given by \cite{Barrow_2020tzx}
\begin{equation}
\label{barrowentropy}
S = \left(S_{BH}\right)^{1+\frac{\delta}{2}},
\end{equation}
where \(S_{BH}\) is the standard Bekenstein–Hawking entropy and $0 \le \delta \le 1$ is a Barrow parameter quantifying the deviation from a smooth spacetime. In this formulation, \(\delta = 0\) recovers the standard Bekenstein–Hawking entropy, while \(\delta = 1\) corresponds to the maximum fractal influence on the black hole surface.
\\

Barrow entropy of black holes, Eq.~\eqref{barrowentropy}, is a direct result of Barrow's model, which analogizes spacetime foam with the Koch snowflake. Indeed, by constructing a hierarchy of ever-smaller spherical bumps on the black hole event horizon, the spacetime foam is endowed with an effective fractal dimension
\begin{equation}
d_{\mathrm{eff}} = 2 + \delta,    
\end{equation}
where \(\delta\) quantifies the deviation from a smooth surface. Since in fractal geometry every fractal system has a dimension exceeding its topological dimension, \(\delta\) is always positive. Moreover, quantum fluctuations in spacetime increase the complexity of its structure at the Planck scale rather than smoothing it out. This increased complexity leads to a greater number of microstates compared to a smooth horizon, i.e., in the absence of spacetime foam. Consequently, the measured entropy exceeds the value proportional to the area.
\\

To incorporate the effects of spacetime structure on charged black holes in Anti-de Sitter spacetime, we adopt the following metric
\begin{equation}
\label{metric}
ds^2 = -f(r) \, dt^2 + \frac{dr^2}{f(r)} + r^2 \, d\Omega^2,
\end{equation}
where the metric function is defined as
\begin{equation}
\label{f}
f(r) = 1 - \frac{2M}{r} + \frac{Q^2}{r^2} + \frac{r^2}{\ell^2},
\end{equation}
where \(M\), \(Q\), and \(\ell\) represent the black hole's mass, electric charge, and the AdS radius, respectively. The AdS radius is related to the cosmological constant by $\Lambda = -3/\ell^2$. The black hole's mass is obtained by solving \(f(r_+) = 0\), where \(r_+\) denotes the event horizon radius
\begin{equation}
\label{M}
M = \frac{r_+}{2}\left(1 + \frac{Q^2}{r_+^2} + \frac{r_+^2}{\ell^2}\right).
\end{equation}
Furthermore, the Barrow entropy of these black holes, expressed in terms of the event horizon radius, is given by
\begin{equation}
\label{Sb}
S = \left(\pi\, r_+^2\right)^{1+\frac{\delta}{2}},
\end{equation}
To explore the impact of the quasi-fractal structure spacetime  on black hole thermodynamics, we work within the extended phase space thermodynamics, treating the cosmological constant as a thermodynamic variable related to the pressure \cite{Kastor_2009wy}
\begin{equation}
P = -\frac{\Lambda}{8\pi} = \frac{3}{8\pi\, \ell^2}.
\end{equation}
The first law of thermodynamics for charged AdS black holes in the extended phase space is given by
\begin{equation}
\label{FL}
dM = T\, dS + \Phi\, dQ + V\, dP.
\end{equation}
From this law, we determine additional thermodynamic quantities. The thermodynamic volume is defined as
\begin{equation}
\label{Vo}
V = \left(\frac{\partial M}{\partial P}\right)_{S,Q} = \frac{4}{3}\pi\, r_+^3,
\end{equation}
and the electric potential at the event horizon is
\begin{equation}
\label{Phi}
\Phi = \frac{Q}{r_+}.
\end{equation}

To examine the influence of the quasi-fractal structure of spacetime on the thermal evolution and emergent thermodynamic behavior, we derive the modified black hole temperature from the first law, Eq.~\eqref{FL}, as follows
\begin{equation}
\label{Tb}
T = \left(\frac{\partial M}{\partial S}\right)_{Q,P} = \frac{8\pi P\, r_+^4 + r_+^2 - Q^2}{2\pi^{1+\frac{\delta}{2}}(\delta+2)\,r_+^{3+\delta}}.
\end{equation}
Furthermore, we construct a corrected Smarr relation by expressing the black hole mass in terms of its thermodynamic variables. We obtain the following expression
\begin{equation}
\label{SR}
M = \left(2+\delta\right) T S + \Phi Q - 2 P V,
\end{equation}
where, the term \(\delta\, T S\) represents the impact of the quasi-fractal structure.  

\section{Thermodynamics}
\label{SS2}
In this section, we analyze emergent thermodynamic phenomena and examine the influence of the quasi-fractal structure on black hole thermodynamics, including phase transitions and stability. Starting from Eq.~\eqref{Tb}, we construct the equation of state for charged AdS black holes as
\begin{equation}
\label{EoS}
P = \frac{\left(1+\delta\right)^2 \left(\pi r_+^2\right)^{\delta/2} T}{4r_+} + \frac{Q^2}{8\pi r_+^4} - \frac{1}{8\pi r_+^2}.
\end{equation}
For \(\delta = 0\), this equation reduces to the standard equation of state for a charged AdS black hole without spacetime foam corrections. Moreover, setting \(Q=0\) in  Eq.~\eqref{EoS} yields the equation of state for a Schwarzschild-AdS black hole that incorporates the quasi-fractal structure.  This equation of state illustrates the impact of the natural structure of spacetime on black hole behavior. To uncover this impact, we determine the critical point by solving the following equations
\begin{equation}
\frac{\partial P}{\partial r_+} = 0,\quad \qquad \frac{\partial^2 P}{\partial r_+^2} = 0.
\end{equation}
Thus, we determine the critical quantities as
\begin{equation}
\label{pc}
r_c = \sqrt{\frac{2(\delta+3)}{\delta+1}}\, Q.
\end{equation}
\begin{equation}
\label{rc}
P_c= \frac{(\delta +1)^2}{32 \pi  (1- \delta ) (\delta +3) Q^2},
\end{equation}
and
\begin{equation}
\label{tc}
T_c = \frac{2^{\frac{1-\delta}{2}}\, (1+\delta)^{\frac{1+\delta}{2}}}{\pi^{\frac{2+\delta}{2}}\, (\delta+3)\, (\delta^2+\delta-2)\, \left[(\delta+3)Q^2\right]^{\frac{1+\delta}{2}}}.
\end{equation}
These expressions indicate that the critical phenomena are influenced by the spacetime foam. Consequently, the critical ratio can be written as
\begin{equation}
\rho = \frac{2 P_c\, r_c}{T_c} = \frac{2^{\frac{\delta}{2}-4}\, \pi^{\frac{\delta}{2}}\, (1+\delta)^{\frac{3-\delta}{2}}\, (2+\delta)\, \left(Q^2(3+\delta)\right)^{\frac{1+\delta}{2}}\, \sqrt{\frac{3+\delta}{1+\delta}}}{Q}.
\end{equation}
The critical ratio, when compared with the standard value of \(\rho = 3/8\) for a Van der Waals fluid, demonstrates the modifications induced by the quasi-fractal structure; notably, the standard value is recovered when \(\delta = 0\).
\\

\begin{figure}[htp]
	\centering
	\includegraphics[scale=0.6]{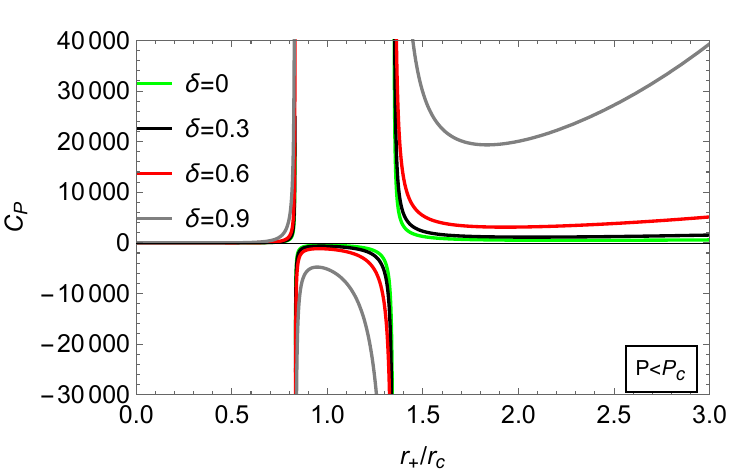}
	\includegraphics[scale=0.6]{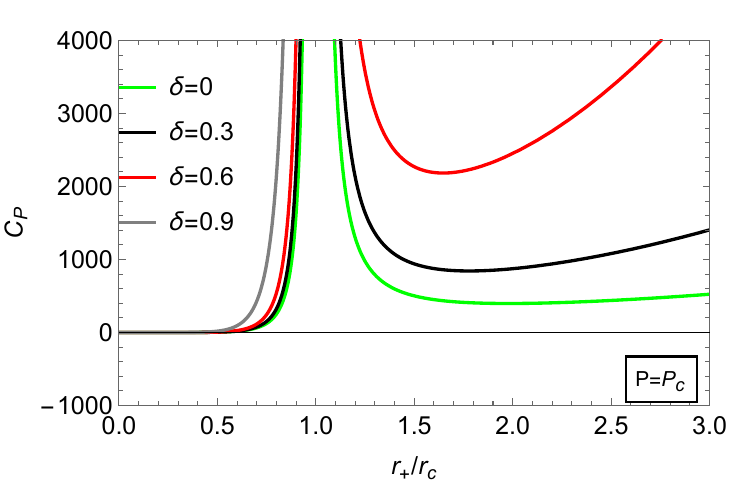}
	\includegraphics[scale=0.6]{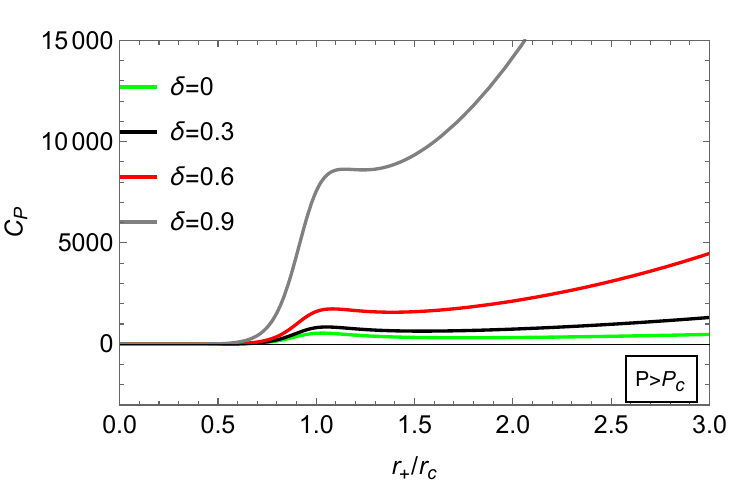}
	\caption{Heat capacity curves for various values of the Barrow parameter and pressure.}
	\label{HC}
\end{figure}

To assess the stability of black holes and further elucidate the impact of the quasi-fractal structure, we compute the heat capacity at constant pressure and charge
\begin{equation}
\label{CP}
C_p = T \left(\frac{\partial S}{\partial T}\right)_{P,Q} 
= \frac{\pi^{1+\frac{\delta}{2}} (\delta+2) r^{\delta+2} \left(8\pi P r^4 - Q^2 + r^2\right)}
{(\delta+3)Q^2 - r^2 \left[(\delta+1) + 8\pi P (\delta-1)r^2\right]}.
\end{equation}
This equation shows that the quasi-fractal structure influences not only the temperature and entropy but also the stability of black holes, underscoring the significant role of the inherent spacetime structure in their thermodynamic behavior.
\begin{figure}[htp]
	\centering
	\includegraphics[scale=0.6]{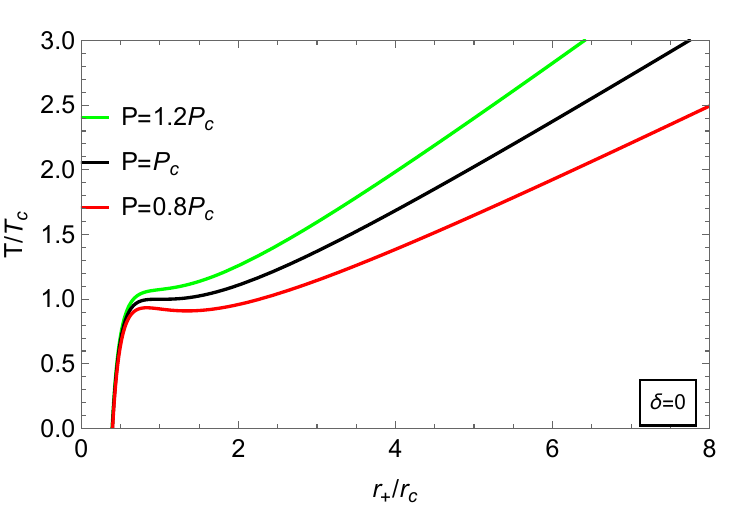}
	\includegraphics[scale=0.6]{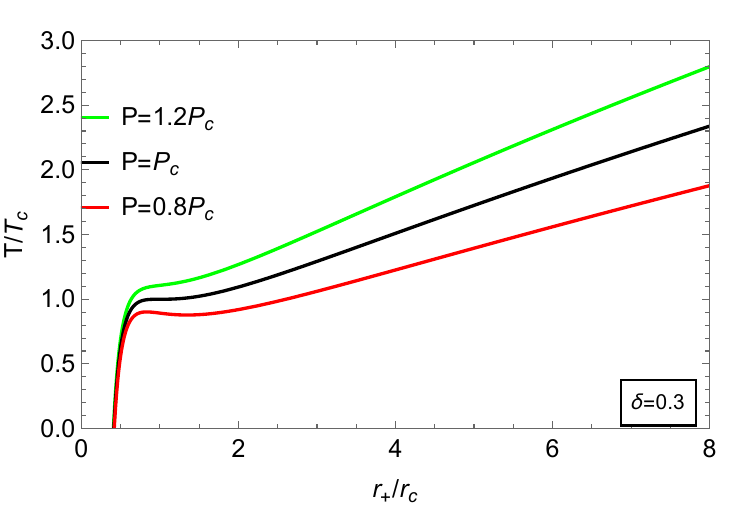}
	\includegraphics[scale=0.6]{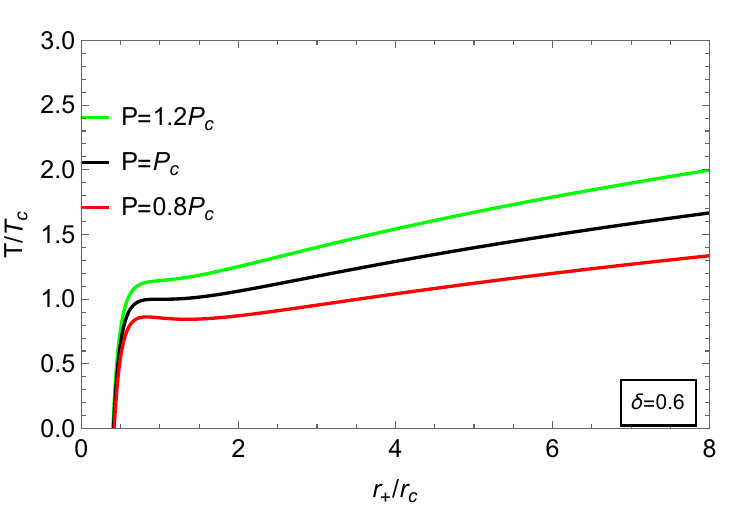}
	\includegraphics[scale=0.6]{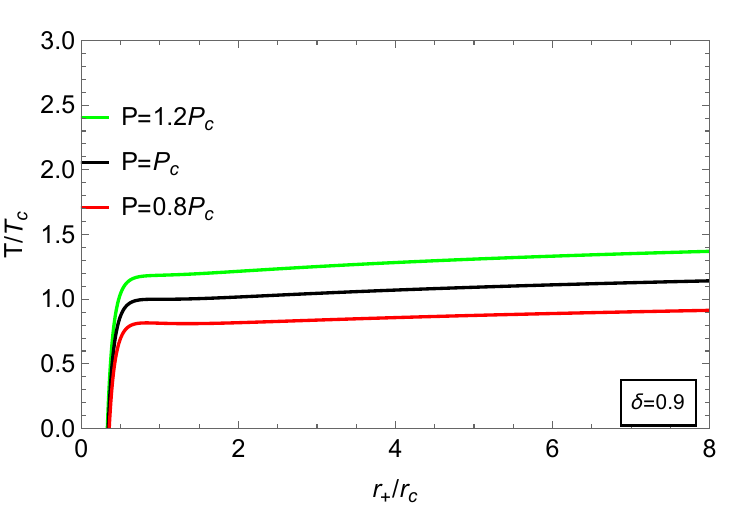}
	\caption{\(T\)-\(r_+\) curves in isobaric processes for various values of pressure and Barrow parameter.}
	\label{T11}
\end{figure}
Fig. \ref{T11} llustrates the thermal evolution of black holes in terms of the event horizon for different values of the Barrow parameter and thermodynamic pressure. We observe that the thermodynamic behavior of black holes is affected by both the pressure and the structure of spacetime. In our analysis, we begin by examining the impact of the thermodynamic pressure. In the first case, when \(P < P_c\) during isothermal processes, we observe first-order phase transitions between small, medium, and large black holes. These phase transitions occur at the event horizons \(r_1\) and \(r_2\), which we determine by solving the following equation 
\begin{equation}
\left(\dfrac{\partial T}{\partial S} \right)_{P, Q}=0,
\end{equation}
where we find two event horizons expressed as 
\begin{equation}
r_1=\sqrt{\frac{1 + \delta -\sqrt{(1 + \delta)^2 - 32 P \pi Q^2 (1 - \delta) (3 + \delta)}}{16 P (1 - \delta) \pi}},
\end{equation}
and
\begin{equation}
r_2=\sqrt{\frac{1 + \delta + \sqrt{(1 + \delta)^2 - 32 P \pi Q^2 (1 - \delta) (3 + \delta)}}{16 P (1 - \delta) \pi}}.
\end{equation}
 Fig. \ref{HC} illustrates the evolution of the heat capacity. It shows that small and large black holes are stable, as they exhibit a positive heat capacity. In contrast, medium black holes are unstable because of their negative heat capacity. In the second case, a second-order phase transition occurs between small and large black holes at \(P = P_c\). In the final case, no phase transition occurs when \(P > P_c\) in isobaric processes, and the heat capacity remains positive. This implies that black holes are always stable and that their behavior mimics that of a Van der Waals fluid.
\\

\begin{figure}[htp]
	\centering
	\includegraphics[scale=0.6]{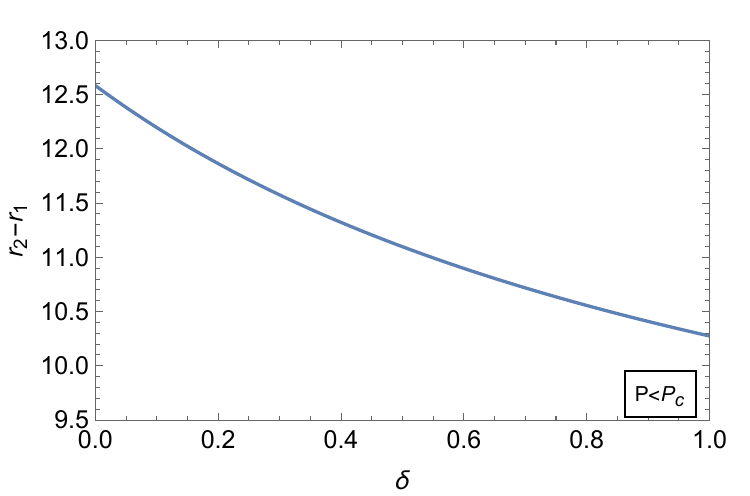}
	\caption{Variation of $r_2 - r_1$ in terms of $\delta$ in isobaric processes with $P < P_c$.}
	\label{Dr}
\end{figure}

We observe the impact of the quasi-fractal structure of spacetime on the thermodynamic behavior of black holes by analyzing how variations in the Barrow parameter affect their thermodynamic properties. An increase in the Barrow parameter leads to a corresponding increase in the critical pressure and temperature. In particular, when the Barrow parameter reaches its maximum value, both the critical pressure, Eq.~\eqref{rc}, and the critical temperature, Eq.~\eqref{tc}, diverge, which in turn alters the conditions for phase transitions.

Furthermore, the thermodynamic behavior of black holes in relation to the quasi-fractal structure depends on the entropy. For small event horizons, i.e., low entropy, the Barrow parameter does not significantly influence the thermodynamic characteristics, as illustrated in Fig.~\ref{T11}. In contrast, black holes with medium and large event horizons exhibit pronounced effects: This is also evident in the heat capacity curves, Fig.~\ref{HC}, where the influence of the quasi-fractal structure becomes apparent only for larger event horizons or higher entropy. For example, the oscillatory features in the \(T\)-\(r_+\) curves corresponding to medium black holes, Fig.~\ref{T11} diminish as the Barrow parameter increases.

It is indeed known that smaller black hole horizons tend to exhibit larger relative quantum fluctuations compared to larger horizons. However, this observation does not contradict our results. Our study focuses on the phenomenological impact of spacetime foam on the thermodynamic behavior of black holes. While small black holes experience larger quantum fluctuations, their thermodynamic quantities, such as temperature, heat capacity, and entropy, are relatively small. Consequently, the overall impact of spacetime foam on these small black holes is limited. In contrast, large black holes possess significantly higher values of these thermodynamic quantities, making the phenomenological effects of spacetime foam more pronounced, despite their smaller relative quantum fluctuations.
In other words, spacetime foam influences are present at all scales but become thermodynamically effective only when the underlying variables reach sufficiently large values.

Additionally, Fig.~\ref{Dr} demonstrates that the difference between the event horizons at which first-order phase transitions occur decreases with increasing Barrow parameter. This suggests that the quasi-fractal structure affects the stability of black holes. Indeed, as the Barrow parameter increases, the region corresponding to medium-sized black holes, i.e. the instability region, shrinks. For black holes with large entropy, the temperature or pressure evolves linearly. However, as the Barrow parameter increases, the slope of this linear behavior decreases, eventually becoming nearly constant as the Barrow parameter approaches its maximum value, as shown in Fig.~\ref{T11}.

We thus conclude that the presence of spacetime foam slows their evaporation of large black holes, implying that the fractal nature of spacetime acts as a resistance to their evaporation.

\section{Joule-Thomson expansion}
\label{SS3}
After analyzing the influence of the quasi‐fractal structure on the thermodynamic behavior of charged AdS black holes in isobaric and isothermal processes, we now turn to its effect in isenthalpic processes, specifically the Joule–Thomson expansion. In this context, the Joule–Thomson expansion of AdS black holes captures how the Anti–de Sitter background alters their thermal evolution under constant enthalpy. We investigate the thermal regimes of these black holes as the pressure decreases and assess how spacetime foam modifies these regimes. To this end, we examine the variation of temperature with respect to pressure at fixed black hole mass, quantified by the Joule–Thomson coefficient \(\mu\) \cite{Okcu_2017qgo}
\begin{equation}
\label{JT}
\mu \;=\;\Bigl(\frac{\partial T}{\partial P}\Bigr)_{M}
\;=\;\frac{1}{C_{P}}\Bigl[T\,\Bigl(\frac{\partial V}{\partial T}\Bigr)_{P}-V\Bigr].
\end{equation}
The pressure in this process always decreases. We determine the regions of cooling and heating by the sign of the Joule-Thomson coefficient. If the coefficient is positive, the temperature decreases, indicating that a cooling region occurs. Conversely, if it is negative, the temperature increases, indicating that a heating region occurs. When the Joule-Thomson coefficient is zero, an inversion occurs between these two phases. We express the inversion temperature $T_i$ as follows
\begin{equation}
\label{Ti}
T_i = V\left(\frac{\partial V}{\partial T}\right)_P 
= \frac{(\delta+3) Q^2 - r_+^2\left[\delta + 8\pi (\delta-1) r_+^2\,P_i + 1\right]}{6(\delta+2)\pi^{1+\frac{\delta}{2}}r_+^{\delta+3}}
\end{equation} 
where $P_i$ represents the inversion pressure. From Eq.~\eqref{Ti}, we see that the inversion temperature depends on the Barrow parameter, indicating that spacetime foam influences the boundary between the cooling and heating regimes. We can also express the inversion temperature as a function of the inversion pressure using the equation of state given by Eq. \eqref{Tb}, as follows
\begin{equation}
\label{EoS2}
T_i = \frac{8\pi P_i\, r_+^4 - Q^2 + r_+^2}{2\,\pi^{1+\frac{\delta}{2}}\, (\delta+2)\, r_+^{3+\delta}}.
\end{equation}
By utilizing the equality between Eq. \eqref{Ti} and Eq. \eqref{EoS2}, we find an equation for the event horizon
\begin{equation}
\label{Pi}
(\delta +6) Q^2-r_+^2 \left(\delta +8 \pi  (\delta +2) P_i r_+^2+4\right)=0.
\end{equation}
We obtain the expression for the event horizon corresponding to the inversion temperature by solving Eq. \eqref{Pi}
\begin{equation}
r_+ = \frac{\sqrt{\sqrt{(\delta +4)^2+32 \pi  (\delta +2) (\delta +6) P_i Q^2}-4-\delta }}{4 \sqrt{\pi\,(\delta +2) P_i }}.
\end{equation}   
We substitute the expression for the event horizon of the inversion temperature in Eq. \eqref{EoS2}, the inversion temperature is given by
\begin{equation}
\label{Tii}
T_i =  \frac{\left(\delta +32 \pi  (\delta +2) P_i Q^2-\sqrt{(\delta +4)^2+32 \pi  (\delta +2) (\delta +6) P_i Q^2}+4\right)\, \Psi}{\sqrt{\pi } (\delta +2)^3 P_i},
\end{equation}
where 
\begin{equation}
\Psi = 4^{\delta +1} \left(\frac{(\delta +2) P_i}{\sqrt{(\delta +4)^2+32 \pi  (\delta +2) (\delta +6) P_i Q^2}-4 -\delta}\right)^{\frac{1}{2} (\delta +3)}.
\end{equation}
\begin{figure}[htp]
	\centering
	\includegraphics[scale=0.8]{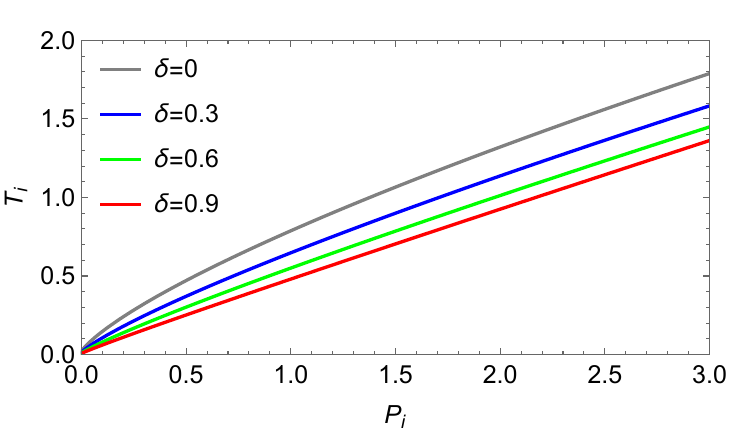}		
	\caption{Inversion temperature curves for different values of the Barrow parameter. }
	\label{TiPi}
\end{figure}
\\

Fig.~\ref{TiPi} represents the curve of the inversion temperature \(T_i\) as a function of the inversion pressure \(P_i\) for different values of the Barrow parameter \(\delta\). We observe that the slope of the inversion curves decreases as \(\delta\) increases. This indicates that the structure of spacetime affects the conditions for the phase transition between cooling and heating regimes: a greater complexity of spacetime (larger \(\delta\)) lowers the inversion temperature. Thus, in addition to the influence of the anti–de Sitter background (via the pressure), there is another factor governing the thermal evolution of black holes, namely the structure of spacetime at the Planck scale. Since spacetime foam originates from quantum gravity effects, this study provides a new perspective to uncover the microstructure of spacetime, the nature and the impact of quantum gravity through emergent thermodynamic phenomena in the Joule–Thomson expansion.

We can determine the minimal inversion temperature using Eq. \eqref{Tii} with $P_i \to 0$. We find
\begin{equation}
T_i^{min}=\frac{\left(\delta +4\right)^{\frac{1}{2}\left(1+\delta \right)}}{\pi ^{1+\frac{\delta }{2}}\,\left(\delta ^2+8 \delta +12\right) \left( \delta +6\right)^{\frac{1}{2}\left(1+\delta \right) } }.
\end{equation}
\begin{figure}[htp]
	\centering
	\includegraphics[scale=0.8]{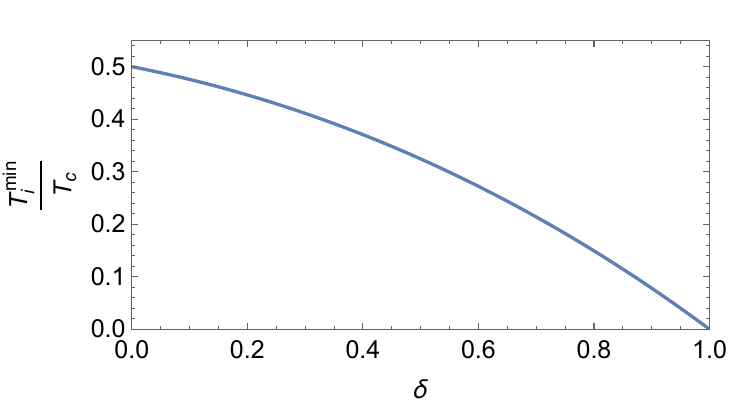}		
	\caption{Curve of the ratio between the minimum inversion temperature and critical temperature, $T_i/T_c$, in terms of $\delta$. }
	\label{Tdelta}
\end{figure}
\\

Fig. \ref{Tdelta} shows the effect of the quasi-fractal structure on the minimum inversion temperature. It is observed that the ratio between minimum inversion and critical temperatures decreases with the growing influence of quasi-fractal structure. In case $\delta = 0$, we recover the ratio between minimum inversion and critical temperatures of regular RN-AdS black holes \cite{Okcu_2017qgo}, $T_i^{min}/T_c = 1/2 $, i.e., without considering the influence of quantum gravity on the black hole surface.
\section{  DENSITY OF MICROSTATES ANS EVAPORATION TIME}
\label{ss11}
In this section, we study the impact of spacetime foam on the density of microstates, the evaporation time of black holes, asymptotic observers, and Unruh detectors, through the use of the Barrow model.

\subsection{ Density of microstates}
Spacetime foam represents the complexity and quantum fluctuations of spacetime at the Planck scale, which also affect the event horizon of black holes. In the Barrow model, the black hole entropy depends both on the horizon area and on these quantum fluctuations, characterized by the Barrow parameter \(\delta\). To study the impact of spacetime structure on the microstate density, we employ the relation between the microstate density, \(\rho\), and the corresponding entropy, \(S\) \cite{TP1,TP2}
\begin{equation}
    S = \ln \rho \,.
\end{equation}
Taking into account the spacetime foam in consideration and the corrected Barrow entropy, the microstate density \(\rho_B\) is expressed as
\begin{equation}
    \rho_B = \exp\!\bigl[(S_{BH})^{1 + \delta/2}\bigr] \,,
\end{equation}
where \(S_{BH}\) is the Bekenstein–Hawking entropy. At \(\delta = 0\), we recover the microstate density without spacetime foam consideration. We conclude that the microstate density with spacetime foam consideration is greater than the microstate density without spacetime foam consideration, so the quantum fluctuations in spacetime at the Planck scale increase the microstate density.
 
\subsection{Black hole evaporation}
Black holes are vaporized by the emission of Hawking radiation \cite{Hawking_1975vcx}, through which they lose mass. This process continues until the emission of radiation stops due to either the complete disappearance of the black hole or the attainment of an extremal black hole state. We study the effect of spacetime foam on the evaporation and lifetime of black holes using the Barrow model. The black hole evaporation can be expressed through blackbody radiation within the Barrow model
\begin{equation}
    \frac{dM}{dt} = -\sigma A T_B^4,
\end{equation}
where \(\sigma\) is the Stefan–Boltzmann constant, \(A = 4\pi r_+^2\) is the black hole surface area, and \(T_B\) is the Barrow temperature (Eq.~\eqref{Tb}). To simplify the calculation, we set \(Q=0\) and \(P=0\), so that the event horizon becomes \(r_+ = 2M\). The Barrow temperature in terms of the black hole mass is then
\begin{equation}
    T_B = \frac{1}{2\,\pi^{1 + \delta/2}(2 + \delta)\,(2M)^{1 + \delta}}.
\end{equation}
We calculate the lifetime of the black hole as
   \begin{equation}
    t_{ev} = \int_{0}^{M_0} -\frac{dM}{\sigma A T_B^4} \propto M_0^{3 + 4\delta},
\end{equation}
where \(M_0\) is the initial mass of the black hole. From this result, we observe that the lifetime of a black hole with spacetime foam consideration is longer than without it. We conclude that spacetime foam, or quantum fluctuations at the Planck scale, acts as a resistance to black hole evaporation. This result generalizes the finding in Section \ref{SS2}, that is, spacetime foam acts as a resistance to black hole phase transitions and evaporation.

\subsection{Asymptotic, local, and Unruh temperatures }

We study the black hole temperature as seen by asymptotic and local observers, taking into account the quasi-fractal spacetime corrections.  Tolman’s law  reads \cite{tb11,tb12}
\begin{equation}
  T_{\rm loc}(r)\,\sqrt{f(r)} \;=\; T_\infty\,,
\end{equation}
where \(T_\infty\) is the temperature measured at spatial infinity.  In our setup \(T_\infty\) coincides with the fractal‐corrected Hawking temperature \(T_B\) (see Eq.\ \eqref{Tb}).  Hence the local temperature is
\begin{equation}
  T_{\rm loc}(r)=\frac{T_B}{\sqrt{f(r)}}\,. 
\end{equation}
As \(r\to r_+\), \(f(r)\to0\) and \(T_{\rm loc}(r)\to\infty\), reflecting the infinite blueshift experienced by a static observer near the horizon.  This divergence is the same phenomenon encountered in the absence of spacetime foam corrections. \\

By contrast, the Unruh temperature depends only on the observer’s proper acceleration \(a(r)\) and is unaffected by the quasi-fractal structure of spacetime.  A static observer at radius \(r\) has \cite{tb13,tb14}
\begin{equation}
  a(r)
  = \frac{\kappa}{\sqrt{f(r)}}
  = \frac{f'(r_+)}{2\,\sqrt{f(r)}}\,,
\end{equation}
where \(\kappa = f'(r_+)/2\) is the surface gravity.  The corresponding Unruh temperature is 
\begin{equation}
  T_U(r)
  = \frac{a(r)}{2\pi }
  = \frac{f'(r_+)}{4 \pi \,\sqrt{f(r)}}\,.
\end{equation}
Since the Unruh temperature \(T_U(r)\) depends solely on the local proper acceleration \(a(r)\), it remains unaffected by quasi-fractal corrections.

\section{Conclusion and Discussion}
\label{SS4}
In this paper, we have explored the effects of quantum gravity on the thermodynamics of charged AdS black holes by incorporating a quasi-fractal structure of spacetime, as described by the Barrow model, into the extended phase space formalism. Our results demonstrate that the quasi-fractal correction to black hole entropy leads to substantial modifications in the thermodynamic behavior, including a revised equation of state and critical parameters. Specifically, we find that an increase in the Barrow parameter results in higher critical pressure and temperature, with these parameters diverging as the quasi-fractal deformation reaches its upper limit. This behavior highlights the significant influence of spacetime foam on black hole phase transitions.
\\

Our analysis further indicates that the spacetime foam effects are predominantly observable in black holes with medium to large event horizons, where both stability and phase structure are significantly affected. The examination of heat capacity curves confirms that, while small black holes (with low entropy) remain largely insensitive to quasi-fractal corrections, the thermodynamic properties of larger black holes are considerably altered. Moreover, the study of the Joule–Thomson expansion reveals that the quasi-fractal structure modifies the inversion curves and the phase transition between cooling and heating regimes. We also study the impact of spacetime foam on black hole evaporation, microstate density, and the observation of black hole temperature. We find that spacetime foam prolongs the evaporation lifetimes, thereby acting as a resistance to black hole evaporation. Additionally, it increases the density of microstates because of the enhanced complexity and quantum fluctuations at the black hole surface. Although local observers experience the expected Tolman blueshift, Unruh temperatures remain unchanged.
\\

In this work, we study the effects of quantum gravity, represented by the quasi-fractal structure of spacetime according to the Barrow model, on the thermodynamics of charged black holes. Thermodynamics, as a phenomenological science, focuses on macroscopic effects rather than microscopic details. Our analysis reveals that the Barrow parameter significantly influences the behavior of large and medium black holes, which are characterized by high entropy and high temperature, while it has a negligible effect on small black holes with low entropy and low temperature. These findings are consistent with recent studies on the effects of quantum gravity on blackbody radiation \cite{BBR} and with previous work on charged black holes \cite{45}.
\\

Future research could expand this work by studying the influence of spacetime structure on other types of black holes, such as rotating black holes. It could also refine the geometric models of spacetime foam and explore their broader implications for cosmology and particle physics.

\section*{Acknowledgments}

Y. Ladghami would like to express gratitude for the support received from the "PhD-Associate Scholarship – PASS" grant (number 42 UMP2023) provided by the National Center for Scientific and Technical Research in Morocco.

\end{document}